# Phase diagram of the vortex state in an amorphous $Re_6Zr$ thin film exhibiting inverse melting

Pritam Das[*], Subhamita Sengupta[*], Anjan Jana, Rishabh Duhan, Sulagna Dutta, Arghya Dutta, John Jesudasan, Vivas Bagwe and Pratap Raychaudhuri[†]

*Tata Institute of Fundamental Research, Homi Bhabha Rd, Mumbai 400005, India.*

In Type II superconductors, the vortex lattice can exhibit "inverse melting," transitioning from a liquid to a crystalline solid as temperature increases. While recently observed via scanning tunneling microscopy in a 20 nm thick amorphous $Re_6Zr$ thin film, this work investigates the corresponding d.c. transport and low-frequency magnetic screening responses. By identifying distinct signatures of these transitions and integrating scanning tunneling spectroscopy imaging, we construct a comprehensive vortex-state phase diagram in the magnetic field-temperature parameter space. Furthermore, we demonstrate that inverse melting is thickness-dependent: a 5 nm film retains an inhomogeneous liquid state, while a 50 nm film maintains a crystalline solid structure except near the upper critical field.

[*] These authors contributed equally.

[†] E-mail: pratap@tifr.res.in

# I. Introduction

In Type II superconductors, above the lower critical field, $H_{c1}$, magnetic field penetrates the superconductor in the form of quantized flux tubes, called vortices[1]. In conventional bulk superconductors these vortices either form a hexagonal vortex lattice[2], the so-called Abrikosov lattice, or settle in a disordered configuration, the vortex glass[3,4,5,6,7]. In bulk single crystals, vortex liquid states, where the vortices are mobile under the influence of thermal or quantum fluctuations, are rare. In contrast, vortex liquid states are readily realized in thin films[8,9,10,11]. Here, the enhanced role of thermal and quantum fluctuations in the 2-dimensional (2D) vortex lattice produces vortex liquid states that can span a significant fraction of the magnetic field-temperature (*H*-*T*) parameter space. Since these vortex liquid states produce stronger dissipation in the presence of a.c. and d.c. transport current, understanding the evolution of these states as a function of temperature[12,13,14], magnetic field and sample thickness is important both from a fundamental perspective as well as from the point of view of applications in superconducting electronics.

In recent years, real space imaging of the vortex state using scanning tunneling spectroscopy[15] (STS) has provided unprecedented insights into the vortex liquid states in superconducting thin films[9,10,11,16]. A variety of vortex liquid states, such as states with hexatic and smectic order, and inhomogeneous vortex liquid states where vortices move through a disordered network of motion paths have been unearthed from these studies. More recently, STS measurements on a 20 nm amorphous $Re_6Zr$ (*a*-ReZr) film revealed the existence of inverse melting[17], a phenomenon theoretically predicted in Type II superconductors[18,19,20,21,22], but rarely observed. In general, inverse melting refers to a situation where a liquid transforms into an ordered solid with increase in temperature[23,24]. In Type II superconductors, inverse melting occurs when a low-temperature vortex liquid freezes into a crystalline solid as temperature rises[25], before eventually remelting into a liquid at even higher temperatures. The same sequence of transformation is also predicted at low temperatures as a function of magnetic field. In *a*-ReZr, the liquid-solid-liquid re-entrant transformation was observed both as a function of temperature at low fields and as a function of magnetic field at low temperature[17]. While this observation provides validation of a long-standing theoretical prediction, one outstanding puzzle is the range of magnetic field over which the low field vortex liquid is observed. While most theories predict the low-field vortex liquid state to be restricted to magnetic fields very close to the lower critical field[22] $H_{c1}$, here the vortex liquid appears to extend up to much higher fields. On the other hand, most of these predictions are for 3-dimensional superconductors in the absence of pinning, so it is important to understand how pinning affects these transformations. STS measurements suffer from two limitations that prevent the accurate determination of the boundaries between various phases in the *H*-*T* plane from STS images alone. First, STS measurements require a continuous smooth area on the film surface without the presence of defects or particulates, that often limits vortex lattice imaging over a relatively small area containing a hundred to a few hundred vortices. This makes it difficult to unambiguously assign the thermodynamic phases close to the phase boundaries. Secondly, with increase in temperature STS imaging loses contrast due to thermal broadening of the superconducting density of states, making it difficult to capture phase transitions that happen close to the superconducting transition temperature, $T_c$. This calls for a detailed investigation of the phase diagram using probes that measure the global response of the vortex lattice.

In this work, we investigate the detailed phase diagram of the re-entrant phase transformation of the vortex lattice in 20 nm thick *a*-ReZr film by combining STS imaging, low-temperature magnetotransport and low-frequency magnetic shielding response measurement using a two-coil mutual inductance technique[26,27]. We utilize the fact that an order-disorder transition of the vortex lattice causes a non-monotonic variation in the pinning properties of the superconductors[6,28,29], which in turn produces visible anomalies on various properties such as the critical current, resistivity and screening response. We can resolve the inverse melting of the vortex lattice in magnetic fields ranging from 0.5 kOe-25 kOe. This inverse melting line appears to end at $0.5T_c$, suggesting that the low field vortex liquid state is confined below this temperature. The vortex solid state on the other hand melts again at higher fields and this melting line extends up to $0.85T_c$. We also compare these results with two other films: 5 nm thick and 50 nm thick *a*-ReZr film. We observe that the re-entrant transition in the vortex state is observed only for the intermediate thickness of 20 nm. In the 5 nm thick film, the vortices remain in an inhomogeneous vortex liquid state over the entire *H*-*T* parameter space. In the 50 nm thick film, we observe a crystalline vortex solid that melts into a liquid close to $H_{c2}$. These results show that in thin films, the re-entrant transformations of the vortex lattice are caused by a complex interplay of dimensionality and disorder.

## II. Experimental Details

*Sample Details:* The samples consist of a-ReZr thin films[30], grown on $Si/SiO_2$ substrates through pulsed laser deposition from a bulk $Re_6Zr$ target. A 248 nm excimer laser with pulse energy 300 mJ/shot and repetition rate of 10 Hz was tightly focused to achieve energy density of the order of ~100 mJ/mm$^2$ on the target. The deposition was carried out at room temperature in a vacuum of $10^{-7}$ Torr. Film thickness was measured using a stylus profilometer for films thicker than 10 nm and estimated from the number of laser shots at lower thickness. The thickness was further confirmed from cross-sectional scanning electron microscope images on films grown under identical conditions. In this work, we investigate films with three thicknesses: 5 nm, 20 nm and 50 nm. In zero-field all films show sharp superconducting transitions with $T_c$ ~ 4.65 K, 5.8 K and 6.8 K respectively (see **Appendix 1**). For STS measurements, we ensure an uncontaminated surface, by directly transferring the films from the deposition chamber into a vacuum suitcase and transferring the film in the scanning tunneling microscope (STM) without exposure to air.

*Scanning tunneling spectroscopy:* STS measurements were performed in a home-built scanning tunneling microscope[31] operating down to 350 mK and fitted with a 90 kOe superconducting solenoid. We used either a mechanically cut Pt-Ir tip or electrochemically etched W tip which were further processed at low temperature through field-emission on a silver single crystal. The vortex state was imaged by measuring the differential tunneling conductance between the tip and the sample $G(V) = \frac{dI}{dV}$ at a fixed d.c. bias voltage and a tunneling current of 250-400 pA, while rastering the tip on the sample surface. The d.c. bias voltage was kept between 1.3-1.45 meV, which corresponded to the coherence peaks in the superconducting density of states. Since at this bias, the $G(V)$ of the superconductor is larger than in a normal metal, the presence of normal vortex cores manifest as a local minimum in the STS image.

*Transport measurement*: Measurements were performed on films of varying thickness grown under conditions identical to the STM samples. To prevent surface oxidation, the films were capped *in situ* with a 1.5 nm-thick Ge layer. For electrical characterization, the samples were patterned into a standard Hall-bar geometry to enable accurate measurements of resistivity and current density. The Hall bar had a width of ~10 μm, with a voltage probe separation of 1.34 mm (see *inset* Fig. 11(b)). Electrical contacts were made using silver epoxy (EPOTEK E4110-10Z) in combination with gold and copper wires to ensure low contact resistance and mechanical stability. Transport measurements were carried out in a $^3$He cryostat equipped with a superconducting solenoid providing magnetic fields up to 110 kOe, with a base temperature of 280 mK. Resistance versus temperature and current–voltage (I-V) characteristics were measured using a conventional four-probe configuration. A Keithley 6220 precision current source and a Keithley 2182A nanovoltmeter were employed for current biasing and voltage detection, respectively. For determining the superconducting transition temperature $T_c$, resistance versus temperature measurements were performed using a fixed current of 500 nA. The critical current, $I_c$, was determined from I-V characteristics measured using a pulsed current protocol to avoid sample heating. To investigate the peak effect, the measurement current was adjusted keeping it close to $I_c$. All current and voltage leads entering the cryostat were equipped with low-pass RC filters[12] with a cutoff frequency of ~ 340 kHz to eliminate high-frequency noise.

*Measurement of a.c. magnetic screening response:* The ac magnetic response of the films was characterized using a two-coil mutual inductance technique[26,27,32], which is particularly effective for superconducting thin films. In this method, an 8 mm diameter film is mounted between a quadrupolar primary drive coil and a dipolar secondary pickup coil (see Fig. 7(a)), such that the alternating magnetic field produced by the primary coil is partially screened by the superconducting film, thereby modifying the flux coupling between the two coils. This approach provides higher sensitivity than conventional ac susceptibility measurements, where the signal is limited by the small volume of thin films. All experiments were conducted in a $^3$He cryostat fitted with a 110 kOe superconducting solenoid, except for the zero d.c. magnetic field response which was measured in a separate 2 K $^4$He cryostat enclosed in a mu-metal shield to eliminate any stray magnetic field. The real and imaginary parts of the mutual inductance ($M'$and $M''$), describing the inductive and dissipative components of the response were recorded as a function of temperature. During the measurements, the ac current in the primary coil was maintained at 0.1 mA at a frequency of 30 kHz, corresponding to an estimated peak magnetic field of ~ 7 mOe at the sample position.

## III. Results

*Evolution of the vortex lattice from STS:* We first compare the structure of the vortex state from STS images of the vortex state in three films with thickness 5, 20 and 50 nm. In Fig. 1, we show the $G(V)$ maps along with their 2-D Fourier transforms (2DFT), acquired at various fields at 460 mK. For the 50 nm sample, the 2DFT shows 6 Bragg spots corresponding to a hexagonal vortex lattice at all fields up to 80 kOe. Here, the vortex lattice remains in a crystalline solid till 85 kOe. For the 5 nm thick sample, the 2DFT shows a ring at all fields, corresponding to a disordered vortex state. In contrast, for the 20 nm film, we observe a non-monotonic variation. At 1 kOe, the 2DFT shows a ring corresponding to a disordered state. As the magnetic field is increased, 6 Bragg spots appear. For 20 and 25 kOe we observe 6 sharp spots in the 2DFT corresponding to a hexagonal vortex solid. Upon further increase in the magnetic field, the Bragg spots broaden and then abruptly transform into a ring above 50 kOe, reflecting the disordering of the ordered vortex solid state. These observations can be further quantified by calculating the metric for orientational order[33,10], $\Psi_6 = (\frac{1}{N})\sum_{i,j} e^{6i(\phi_i-\phi_j)}$, on the set of local minima observed in each $G(V)$ map. $\phi_i$ and $\phi_j$ are the angle between a fixed direction in the plane and the $i$-th ($j$-th) bond connecting two nearest neighbor minima, $N$ is the total number of bonds and the sum runs over all the bonds in the field of view. For a perfect hexagonal lattice $\Psi_6 = 1$, whereas for a completely disordered lattice $\Psi_6 = 0$. For the 5 nm thick film $\Psi_6 < 0.04$, for all fields showing that the vortices remain completely disordered at all fields. For the 20 nm film, $\Psi_6$ increases from 0.03 at 3 kOe to 0.61 at 20 kOe showing that the vortices gradually order with increase in magnetic field. Above 20 kOe, $\Psi_6$ decreases again before abruptly falling to a value close to zero above 50 kOe. For the 50 nm film, $\Psi_6$ varies between 0.5-0.9 between 5-65 kOe and then gradually decreases up to 85 kOe. This gradual decrease happens due to proliferation of defects in the vortex solid, even though the Bragg spots are visible in the 2DFT up to 85 kOe suggesting that vortices remain in a solid state. In Fig. 2, we show the temperature evolution at representative fields for the 50 nm and 20 nm films. For the 50 nm film, the Bragg spots in the 2DFT remain sharp with increase in temperature and disappear at 5 K at 15 kOe and 4 K at 35 kOe. As we will show later, these temperatures are close to the melting line of the vortex lattice determined from transport. On the other hand, for the 20 nm thick film the temperature evolution is more complex. At 3 kOe, the vortices start from a disordered state at low temperature; the Bragg spots gradually become sharper up to 3 K and again become diffused above this temperature. This non-monotonic variation is a signature of inverse melting. On the other hand, at 20 kOe, where the vortex lattice is already ordered at low temperature, increase in temperature has only a very small effect on the Bragg spots. However, we would like to note that at higher temperatures, the contrast in the STS images becomes poor and prevents us from capturing the melting of the vortex solid state close to $T_c(H)$. For the 5 nm thick film, the 2DFT of the $G(V)$ maps (not shown here) show a ring at all temperatures signifying that the vortex lattice remains in a disordered state. The error bars on $\Psi_6$ was obtained by calculating the maximum variation on 10 or 20 consecutive images at the same location. For the 50 nm film this was done for few representative fields.

While the structure of the vortex state captured from individual STS images shows the re-entrant disorder-order-disorder transformation in the 20 nm thick sample, the preceding analysis does not discriminate between a disordered vortex glass where all vortices are static and a vortex liquid where vortices are mobile. For this, we need to analyze the time evolution by capturing a series of successive $G(V)$ maps over the same area. In ref. 11 and 17 this was shown by analyzing relative intensity of local conductance minima in every image and shift in the position of the local minima in successive images. The central conclusion from that analysis was that the vortices form an inhomogeneous liquid where the vortex motion can happen through two different kinds of hops: (i) complete hops from one shallow local minima to another, which leaves the position of the local minimum unchanged in the next image, and (ii) incomplete hops where the vortex gets trapped in an intermediate potential well, which changes the position of the local minima in its vicinity in the next image. Taken together, these hops produce a disordered network of motion paths along which the vortices move in the inhomogeneous vortex liquid state. Instead of repeating this analysis, here we follow a simpler but more intuitive approach. We add all the raw (unfiltered) images taken successively at a given field and temperature and plot them as a function of temperature/magnetic field. In a vortex solid, the local minima appear at the same location in every image and their intensities build up (whereas the random noise gets cancelled) making the contrast in the summed image sharper. On the other hand, when the vortices move as in a liquid, the local minima change positions due to incomplete hops and the contrast in the summed image decreases. In Fig. 3, we show the evolution of the summed images as well as their 2DFT as a function of temperature and magnetic field for the 20 nm thick film. For the 20 nm film, at 460 mK, no vortices can be resolved in the summed image at 1 kOe and the 2DFT is diffuse. With increase in magnetic field diffuse minima start to appear in some parts of the image and the 2DFT shows a diffuse ring. There is a large variation in the intensity of the local minima and in some regions the minima are completely blurred, showing that some vortices are more mobile than others. This is a hallmark of the inhomogeneous vortex liquid which forms under the combined influence of interaction and pinning[11,17]. Between 15-40 kOe, we observe 6 clear Bragg spots in the 2DFT. From the real space summed image we observe that the state at 20 kOe is the most ordered. At 50 kOe the hexagonal order is completely lost and the 2DFT transforms again into a diffuse ring, as expected for an inhomogeneous vortex liquid. The temperature evolution of summed images at 3 kOe (Fig. 4) also shows a similar behavior where the local minima get progressively more pronounced and ordered in the temperature range 0.46 - 3 K, before they get completely blurred at 4 K. One point to note is that the minima appear more resolved in the high field vortex liquid state compared to the vortex liquid state at low fields. This is because at low fields where vortices are far apart the motion is dominated by incomplete hops[17]. In contrast, at high field where vortices are very close to each other, the motion is dominated by complete hops which leaves the position of the minima unchanged. Similar analysis on the 5 nm thick sample (see **Appendix 2**) shows that the vortices remain in an inhomogeneous vortex liquid state over all fields and temperatures.

*Transport signatures of vortex lattice melting:* We now focus on the magnetotransport properties of the a-ReZr films. First, we focus on the magnetic field variation of $I_c$ at 300 mK. Fig. 5(a) shows the typical I-V characteristics for the 20 nm a-ReZr film. The I-V curves show a significant rounding close to $I_c$, arising from flux creep. Therefore, to determine $I_c$, we extrapolate back the linear flux flow region and find $I_c$ from the intercept on the current axis (also see **Appendix 3**). Fig. 5(b)-(d) show the variation of the critical current density $J_c$ with magnetic field for the 5 nm, 20 nm and 50 nm films respectively. In the 5 nm thick sample, the $J_c$ shows a slow monotonic variation up to 80 kOe. Since the vortices are in a completely disordered state in all fields, we expect pinning to act individually on each vortex, for which we expect the variation to follow the form[34,35], $J_c \propto H^{n-1}\left(1-\frac{H}{H_0}\right)$, where $n\sim 0.5-2$ depending on the pinning mechanism and $H_0 \sim H_{c2}$. We see that $J_c$ follows this variation with $n = 0.87$ and $H_0 = 76$ kOe, which is similar to the values reported in ref. 17. In contrast, in the 20 nm thick sample (Fig. 5(b)), $J_c$ follows a non-monotonic behavior. $J_c$ initially decreases at low fields, then gradually increases, reaching a peak at 60 kOe before decreasing again. This peak behavior, known as the "peak-effect"[28,36,37,38] appears when the vortex lattice transforms from an ordered to a disordered state[39]. An alternative, but simpler way to see the peak-effect is to measure the electrical resistance, $R$, of the superconductor keeping the probing current, $I$, in the creep region close to $I_c$. As can be seen in Fig. 5(b), the peak in $J_c$ manifests as a corresponding dip in $R$ at the same field. While the peak effect is normally associated with the transformation of the vortex lattice from a crystalline to an amorphous state, it has been shown that it can appear at a vortex solid to liquid melting transition, as long as the thermally activated flux flow motion in the liquid is much slower than the movement in the flux flow regime[38]. By comparing the earlier STS data, we associate the peak in $J_c$ with the melting of the vortex solid into the high-field liquid state. In the 50 nm thick film, where the vortex forms an ordered solid, $J_c$ decreases rapidly at low fields followed by a more gradual monotonic decrease at higher fields as expected for a weakly pinned Type II superconductor. We do not see any signature of peak-effect, suggesting that at this temperature, the ordered vortex solid does not melt up to 100 kOe.

To track the melting of the vortex lattice in the $H$-$T$ parameter space, we perform a series of $R$-$T$ measurements in different fields. Fig. 6(a) shows $R$-$T$ curves in different magnetic fields for the 20 nm thick film, measured with a probing current of $500$ nA $\ll I_c$. $H_{c2}$ as a function of temperature is determined from the point at which the resistance is 90% of the normal state resistance, $R_N$. In Fig. 6(b) we show a series of $R$-$T$ curves at 20 kOe measured on the same sample with probing currents $I \sim I_c$. The peak effect appears for $I \geq 9$ $\mu$A. This is the signature of the melting of the vortex solid with temperature. It is important to note that while the peak again becomes less pronounced at higher currents, the position of the resistance minima varies at most by 90 mK between the lowest and highest current, which defines the accuracy with which the melting temperature can be determined. Fig. 6(c) shows the peak effect in the R-T curves at different fields, where each curve, $I$ is adjusted to a value close to $I_c$ at that field such that the peak effect is most pronounced. We observe clear signatures of the peak effect between 10-40 kOe. For fields below 10 kOe the signature of peak effect gets masked by the rapid increase in background resistance, but a weak signature is visible down to 6 kOe. For the 50 nm thick sample (Fig. 6(d)) we can similarly track the melting of the vortex lattice from 10-80 kOe.

*Signature of inverse melting from magnetic shielding measurements:* From the transport data we obtain clear signature of the melting of the ordered vortex solid through the “peak effect”. However, we cannot identify any distinct signature associated with the inverse melting observed at lower temperatures in the 20 nm $a$-ReZr film. This could be due to a combination of factors. It can be seen from STS images that unlike the melting transition which happens rather abruptly over a small magnetic field/temperature range, the transformation from the low-field, low-temperature inhomogeneous liquid state to the vortex solid is gradual. This can be seen both from the variation of $\Psi_6$ in Fig. 1 as well as the 2DFT which shows gradual sharpening of the Bragg spots over a range of magnetic fields or temperature. Consequently, the signature of this transformation probably gets masked in the background temperature variation of $J_c$. However, signatures of this transformation are obtained from a.c. magnetic shielding measurements as shown below.

Fig. 7(b)-(c) show $M'$ and $M''$ measured as a function of temperature in different magnetic fields for the 20 nm thick $a$-ReZr film. In zero field below $T_c$, $M'$ decreases rapidly signaling the onset of diamagnetism; on the other hand, $-M''$ shows the characteristic dissipative peak close to $T_c$ and rapidly goes to zero below the transition. Under the application of magnetic field, the oscillatory motion of vortices causes additional coupling between the primary and the secondary, resulting in a change in both $M'$ and $-M''$. $M'$ monotonically decreases below $T_c(H)$ but its magnitude progressively increases with increasing field. The temperature variation of $-M''$ is more complex. Below, the characteristic dissipative peak at $T_c(H)$, $-M''$ also decreases with decreasing temperature, but closer inspection reveals a more complex feature (Fig. 7(d)). At very low fields, we observe that $-M''$ passes through a minimum or displays

a shoulder before decreasing again. Above 25 kOe, this feature disappears and $-M''$ shows a smooth decrease down to low temperatures. This weak anomaly, observed for $H \leq 25$ kOe is associated with the inverse melting of the vortex lattice as we argue below.

In the presence of a.c. magnetic field, the equation of motion of vortices is qualitatively captured within the Gittleman-Rosenbaum[40,41] phenomenological model, where the motion of each vortex is described by an overdamped harmonic oscillator,

$$\eta \frac{du}{dt} + \alpha u = \Phi_0 J^{ac} e^{i\omega t} = F e^{i\omega t},$$

where, $u$ is the position of the vortex, $\eta$ is the Bardeen Stephen viscous drag per unit length of the vortex, $\alpha$ is the effective restoring force per unit length on the vortex, $\Phi_0 = \frac{h}{2e}$ is the flux quantum, $J^{ac} \ll J_c$ is the induced ac current by the oscillatory magnetic field. In this model $\alpha$ is an effective mean-field parameter that has contribution from both the elementary pinning force constant, $\alpha^{el}$, and intervortex interactions. The in-phase and out-of-phase components of the solution, which mimic $M'$and $-M''$, are given by $u' \propto \left(\frac{1}{\alpha}\right)\frac{1}{1+\omega^2\tau^2} \approx \left(\frac{1}{\alpha}\right)$ and $u'' \propto \left(\frac{\omega\eta}{\alpha^2}\right)\frac{1}{1+\omega^2\tau^2} \approx \left(\frac{\omega\eta}{\alpha^2}\right)$ where $\tau = \eta/\alpha$; the last approximation is valid at kHz frequencies where $\omega\tau \ll 1$. In general, elementary pinning force constant and viscous drag follow the approximate relations[14], $\alpha^{el}(t) \sim \alpha_0(1-t^2)^2$ and $\eta \sim \eta_0 \left(\frac{1-t^2}{1+t^2}\right)$ (where $t = T/T_c(H)$), such that $-M''$ smoothly decreases below the dissipative peak at $T_c(H)$. However, this behavior changes when vortex-vortex interactions are included. In the presence of intervortex interactions, the pinning force acts collectively on the vortices[42], and the effective pinning force depends on the degree of order in the vortex lattice. When the vortex lattice is ordered, the rigidity of the vortex lattice does not allow all vortices to be on pinning centers, and consequently the effective $\alpha$ becomes smaller than the elementary pinning force. Thus, the restoring pinning force gets modulated by an additional factor such that, $\alpha(t) = \alpha^{el}(t)f(t)$, where $f(t)$ depends on the degree of order in the vortex lattice. Since inverse melting, the vortex lattice gets maximally ordered at an intermediate temperature, $f(t)$ has a local minimum at this temperature. This, in turn, produces the shoulder (or weak minima) in $-M''$ that we observe in our experiments. To track this feature as a function of magnetic field and temperature, we compute the derivative $\frac{dM''}{dT}$, where it manifests as a local maximum (Fig. 7(e)-(h)) for fields lower than 25 kOe. In principle, this anomaly should also reflect the temperature variation of $M'$. However, since $M'$ is proportional to $\frac{1}{\alpha}$ instead of $\frac{1}{\alpha^2}$, the feature is much weaker there and gets masked in the background (For more details see **Appendix 4**).

*Vortex phase diagram:* We now construct the vortex phase diagram in the *H-T* parameter space, compiling all the data from STS, transport, and magnetic shielding measurements. Fig. 8(a) shows the phase diagram for the 20 nm thick *a*-ReZr film, where the $H_{c2}$ is determined from *R*-*T* measurements in different fields, the melting line is obtained from the peak effect in magnetotransport, the inverse melting line is obtained from magnetic shielding measurements. Also shown are the points for the corresponding transitions obtained from STS. The microscopic STS data are consistent with bulk measurements, and the small difference is attributed to sample-to-sample variation. To compare, in fig. 8(b)-(c) we show the phase diagrams for the 5 nm and 50 nm thick *a*-ReZr films. For the 5 nm thick film, the vortices form an inhomogeneous vortex liquid over the entire *H*-*T* parameter space up to $H_{c2}$. For the 50 nm thick sample, we observe ordered vortex solid at low fields that melt into a vortex liquid close to $H_{c2}$. This behavior is typical of many weakly pinned Type II superconductors. For all three samples $H_{c2}$-*T* can be fitted with the Werthamer–Helfand–Hohenberg (WHH) theory[43] from which we obtain $H_{c2}(0)$ (see **Appendix 5**).

## IV. Discussion

One central observation in this study is that re-entrant transformations of the vortex lattice are observed only for a film with intermediate thickness whereas very thin or very thick films do not display this phenomenon. This points towards a complex interplay of three factors that determine the vortex state in thin films: vortex-vortex interactions that favor the formation of an ordered state, the random pinning potential that tends to destroy this order, and temperature that promotes thermally activated motion of vortices. The thickness of the film affects the vortices in two different ways. First, in amorphous superconductors, where crystallographic defects are not significant, a major component of the pinning comes from surface defects (for example, surface roughness). Since the surface contribution progressively decreases with increasing thickness, the effective pinning becomes weaker with increasing thickness. Secondly, since the condensation energy of the vortex scales directly with film thickness, the thermally activated motion of the vortices becomes more favorable in thinner samples. The behavior in the thick and thin limit can be understood based on the combination of these two effects. In the 50 nm film, the low pinning and larger thermal activation energy stabilizes the ordered vortex solid over much of the *H*-*T* parameter space. Only at elevated temperatures and magnetic fields close to $H_{c2}$, thermal fluctuations melt this ordered state into a liquid. In contrast, in the 5 nm sample where pinning dominates, the ordered vortex solid is destroyed at all fields. However, in contrast to bulk single crystals where strong pinning transforms the ordered vortex solid into a vortex glass, here the concomitant decrease in the thermal activation energy makes thermally activated vortex motion more favorable and drives the vortices into an inhomogeneous vortex liquid. However, evolution between these two scenarios follows a non-trivial route as can be seen from the 20 nm thick *a*-ReZr films. When pinning is intermediate, the high-field vortex liquid expands at the cost of the ordered vortex solid. However, at the same time a pocket of inhomogeneous vortex liquid also appears at low fields and low temperatures. Naively, one would argue that at low fields where the intervortex separation is large, the pinning energy dominates over the weak vortex-vortex interactions driving the ordered solid into a low-field inhomogeneous vortex liquid. Signature of the same effect is also visible in the 50 nm thick film where the vortex solid appears to get more disordered below 10 kOe as evidenced from the decrease in $\Psi_6$ (Fig. 1). However, in that case, the low pinning and larger thermal activation energy does not allow the destruction of the vortex solid order so that an inhomogeneous vortex liquid state is never realized. However, more interesting is that for the 20 nm thick film, the inverse melting line appears to end at 3 K, leaving a region at higher temperatures where the vortices form an ordered solid even at very low fields. Thus, it appears that there is a range of temperatures at low fields where the interactions dominate over pinning, but thermal fluctuations are not yet strong enough to melt the vortex lattice. At present, we do not have an explanation for this. Another distinctive feature of inverse melting is that the onset of order is gradual, as opposed to a sharp first order melting transition. This can be seen from the gradual increase in $\Psi_6$ (Fig. 1) as one approaches the inverse melting transition. Because quenched random disorder is recognized to broaden first-order transitions[44,45] or induce second-order behavior, the observed broadening of the inverse melting transition reinforces the role of random pinning in stabilizing the low-field vortex liquid state.

## V. Conclusion

Using a combination of STS, magnetotransport and low frequency magnetic shielding response measurements we constructed the phase diagram of the re-entrant transformations in the vortex state in 20 nm *a*-ReZr thin films exhibiting inverse melting of the vortex lattice. While our study shows the existence of inverse melting, the phase diagram is strikingly different from theoretical predictions. In most theoretical calculations in 3D superconductors, the high-field and low-field vortex liquid phases are connected through a nose in the phase boundary[19,20,21]. In contrast, in the 20 nm *a*-ReZr film, the high-field and low-field vortex liquid phases are separated by a vortex solid state in between. Furthermore, the absence of re-entrant transition for much thinner (5 nm) or much thicker (50 nm) films points towards a non-trivial role of pinning and thermal fluctuations on the 2-dimensional vortex lattice. Our results suggest that the complex interplay of thermal fluctuations, intervortex interactions, and random pinning can produce unexpected phenomena that should be addressed in future theoretical studies. It would also be interesting to look for similar re-entrant transformations in analogous systems, such as two-dimensional colloidal crystals, charge density waves, and Skyrmion crystals which share many common features with the two-dimensional vortex system.

### Appendix 1: Superconducting transition in zero field

Fig. 9 shows the temperature dependence of resistance $R(T)$ for the 5 nm, 20 nm, and 50 nm $Re_6Zr$ thin films measured in our $^3$He cryostat at zero field. All three samples display a clear superconducting transition marked by a sharp drop in resistance as the temperature is lowered below the superconducting critical temperature $T_c$. The narrow transition width indicates good sample homogeneity and a relatively uniform superconducting order parameter across the films. We also observe an increase in resistivity with decreasing film thickness confirming the increase in disorder with decreasing thickness.

**Appendix 2: Inhomogeneous vortex liquid state in the 5 nm thick a-ReZr thin film**

In ref. 11, it was shown from an analysis of the motion paths that the vortices in 5 nm thick *a*-ReZr thin films remain in an inhomogeneous liquid state at all temperatures and magnetic fields. In Fig. 10 we show representative summed images at various temperatures and magnetic fields on the same sample along with the 2DFT of the summed images. As expected, the real space images are blurred, and the 2DFT always shows a diffuse ring confirming the same conclusion.

**Appendix 3: Determination of critical current**

The current–voltage ($I$–$V$) characteristics of the films were measured in pulsed mode using rectangular current pulses with a duration of 50 ms and a repetition interval of 5 s to minimize Joule heating at high currents. The $I$-$V$ curves for the 20 nm sample are shown in Fig. 5(a). Fig. 11 (a) and 11 (c) show the $I$–$V$ curves for the 50 nm and 5 nm $Re_6Zr$ samples measured at 300 mK.

In the mixed state, the $I$-$V$ curve has three regimes: The zero resistance below the flux-flow critical current, $I_c$; the flux flow regime for $I > I_c$ where $V = R_{ff}(I - I_c)$, where $R_{ff}$ is the Bardeen–Stephen flux-flow resistance, and the depairing regime at higher currents where the resistance abruptly jumps to its normal state value. At finite temperatures, thermally activated vortex motion leads to vortex creep[46,47,22], which smoothens the sharp transition between the zero-resistance regime and the flux-flow regime. As a result, the $I$–$V$ characteristics show a gradual rounding near the critical current. In such cases, $I_c$ is determined by extrapolating the linear flux-flow regime of the $I$–$V$ curve to its intersection with the current axis. This procedure has been used to determine $I_c$ for the 20 nm and 50 nm samples. Fig. 11 (b) shows the $I$–$V$ curve measured at 300 mK and 30 kOe for the 50 nm sample using this procedure.

In two-dimensional superconducting films, however, the formation of a vortex-slush[48,11] in the inhomogeneous vortex liquid phase complicates this analysis. Since different vortices experience different degrees of pinning, the clear demarcation between the flux flow regime and the depairing regime gets obliterated, and we do not obtain a clear linear flux flow regime. To estimate $I_c$ in this situation, we first evaluate $R_{ff} = R_N \frac{B}{H_{c2}}$, using the measured normal-state resistance $R_N$ and the upper critical field $H_{c2}$. A tangent is then drawn to the $I$–$V$ curve at the point where its slope matches $R_{ff}$, and the critical current $I_c$ is obtained from the intersection of this tangent with the current axis. Fig. 11 (d) illustrates this procedure for the 5 nm sample at 300 mK and 30 kOe.

**Appendix 4: Origin of the anomaly in $M''$ close to inverse melting**

Within the collective pinning theory of Larkin and Ovchinnikov[42] the random pinning forces add up such that the effective restoring force on the vortex lattice is proportional to $\alpha \propto \frac{\langle \alpha^{el} \rangle}{tR_c^2}$ where $R_c$ is a characteristic length scale up to which the vortex lattice maintains its order. In general, $R_c$ decreases with increasing temperature till the melting temperature where it becomes the order of the intervortex separation. However, in a situation where we observe inverse melting, the vortices get ordered with increasing temperature and $R_c$ increases till the inverse melting point. After this $R_c$ decreases again till the ordered vortex solid melts. This produces a non-monotonic temperature variation of $\alpha$ that produces a shoulder or weak minimum in $-M''$.

To qualitatively illustrate this effect, define a function $f(t) = 1 - A\exp\left[-\frac{(t-t_0)^2}{2\sigma^2}\right]$ that incorporates this non-monotonic variation, such that $\alpha(t) = \alpha^{el}(t)f(t)$. Fig. 12(a) shows the variation of $\alpha^{el}(t) = (1-t^2)^2$, $f(t)$ and $\alpha(t)$ with $A = 0.3$, $t_0 = 0.5$, and $\sigma = 0.2$, and $T_c = 6$ K. In Fig. 12(b) and 12(c), we show the temperature variation of $M' \propto \frac{1}{\alpha}$ and $-M'' \propto \frac{\omega\eta}{\alpha^2}$ respectively, using the temperature variation of $\alpha^{el}(t)$ and $\eta(t)$ as given in the main text. For comparison we have also plotted the curves where $\alpha(t) = \alpha^{el}(t)$. We observe that while both $-M''$ and $M'$ show an anomaly due to inverse melting, the anomaly in $-M''$ is much more pronounced due to the reason explained in the main text.

**Appendix 5: Determination of the upper critical field *$H_{c2}(0)$***

In the phase diagrams shown in Fig. 8(a)-(c), $H_{c2}$ is defined as the points in the *H-T* plane where the resistance is 90% of the normal state resistance. To obtain the values of $H_{c2}(0)$ we fit the $H_{c2}$-$T$ points with the WHH model[43] that describes the temperature dependence of $H_{c2}$ in conventional type-II superconductors by incorporating the effects of both orbital pair breaking and spin-related mechanisms. The WHH formalism provides a microscopic framework to account for these effects simultaneously. In this model, the temperature dependence of the upper critical field is obtained by solving the linearized gap equation in the presence of a magnetic field. The WHH equation includes two important parameters: the Maki parameter $\alpha$, which quantifies the relative strength of the Pauli paramagnetic pair-breaking effect compared to the orbital limiting mechanism, and the spin–orbit scattering parameter $\lambda_{so}$, which describes the influence of spin–orbit interactions on superconductivity. The WHH equation can be written in terms of the reduced temperature $t = T/T_c$ and the reduced magnetic field $h$, which is related to the slope of the upper critical field near the superconducting transition temperature:

$$h = \frac{4}{\pi^2} \frac{H_{c2}(T)}{-\left|\frac{dH_{c2}}{dT}\right|_{T=T_c} T_c}$$

The implicit WHH relation used to determine $H_{c2}(T)$ is

$$\ln\left(\frac{1}{t}\right) = \left(\frac{1}{2} + \frac{i\lambda_{so}}{4\gamma}\right)\psi\left(\frac{1}{2} + \frac{h + \lambda_{so}/2 + i\gamma}{2t}\right) + \left(\frac{1}{2} - \frac{i\lambda_{so}}{4\gamma}\right)\psi\left(\frac{1}{2} + \frac{h + \lambda_{so}/2 - i\gamma}{2t}\right) - \psi\left(\frac{1}{2}\right)$$

where $\psi$ is the digamma function and

$$\gamma = \sqrt{\left((\alpha h)^2 - \left(\frac{\lambda_{so}}{2}\right)^2\right.}$$

By fitting the experimentally determined $H_{c2}(T)$ values with this model, the parameters $\alpha$ and $\lambda_{so}$ were obtained. The fitting allows us to quantify the relative contributions of orbital depairing, Pauli paramagnetic limiting, and spin–orbit scattering in determining the upper critical field of the system. The extrapolated value of $H_{c2}(0)$ was obtained from the WHH fit, providing an estimate of the upper critical field at zero temperature. The parameters obtained from the fits for the three films are given in Table 1.

*Acknowledgements:* This work was supported by the Department of Atomic Energy, Government of India. PD performed the transport measurements and analyzed the data. PD, SS and SD performed the two-coil mutual inductance measurement and analyzed the data. RD, AJ, SD and PD performed the STS measurements and PD and RD analyzed the data. AD, JJ and VB provided technical support in thin films growth and measurements. PR conceived the problem, supervised the experiments and analysis and wrote the paper with inputs from all authors.

*Data availability:* All data included in this paper are available from the corresponding author upon reasonable request.

---

[39] The peak effect was not captured in ref. 17 where transport measurement was performed on a 300 μm wide channel instead of the 10 μm wide channel here. Consequently, the critical current was very high and the linear flux flow regimes in the I-V curves could not be captured due to heating.

[40] J. I. Gittleman and B. Rosenblum, Radio-frequency resistance in the mixed state for subcritical currents Phys. Rev. Lett. 16, 734 (1966).

[41] J. I. Gittleman and B. Rosenblum, The Pinning Potential and High-Frequency Studies of Type-II Superconductors, Journal of Applied Physics **39**, 2617 (1968).

[42] A. I. Larkin, Y. N. Ovchinnikov, Pinning in type II superconductors. *J Low Temp Phys* **34**, 409 (1979).

[43] N. R. Werthamer, E. Helfand, and P. C. Hohenberg, Temperature and Purity Dependence of the Superconducting Critical Field, $H_{c2}$. III. Electron Spin and Spin-Orbit Effects, Phys. Rev. 147, 295 (1966).

[44] Y. Imry and M. Yortis, Influence of quenched impurities on first-order phase transitions, Phys. Rev B 19, 3580 (1979).

[45] M. Aizenman and J. Wehr, Rounding Effects of Quenched Randomness on First-Order Phase Transitions, Commun. Math. Phys. 130, 489 (1990).

[46] P. W. Anderson, Theory of flux creep in hard superconductors, Phys. Rev. Lett. 9, 309–311 (1962).

[47] P. W. Anderson and Y. B. Kim, Hard superconductivity: Theory of the motion of Abrikosov flux lines, Rev. Mod. Phys. 36, 39–43 (1964).

[48] Y. Nonomura and X. Hu, Effects of point defects on the phase diagram of vortex states in high-$T_c$ superconductors, Phys. Rev. Lett. 86, 5140–5143 (2001).

**Table 1. Fit parameters from WHH fit for different $Re_6Zr$ film thickness.**

| Thickness | $T_c$ (K) | $\lambda_{so}$ | $\alpha$ | $\frac{dH_{c2}}{dT}$ ( kOe/K) | $H_{c2}(0)$ ( kOe ) |
|---|---|---|---|---|---|
| 5 nm | 5.2 | 0.07 | 0.95 | 156 | 81.50 |
| 20 nm | 6.1 | 0.07 | 0.51 | 169 | 109.15 |
| 50 nm | 6.8 | 0.07 | 0.43 | 181 | 120.65 |

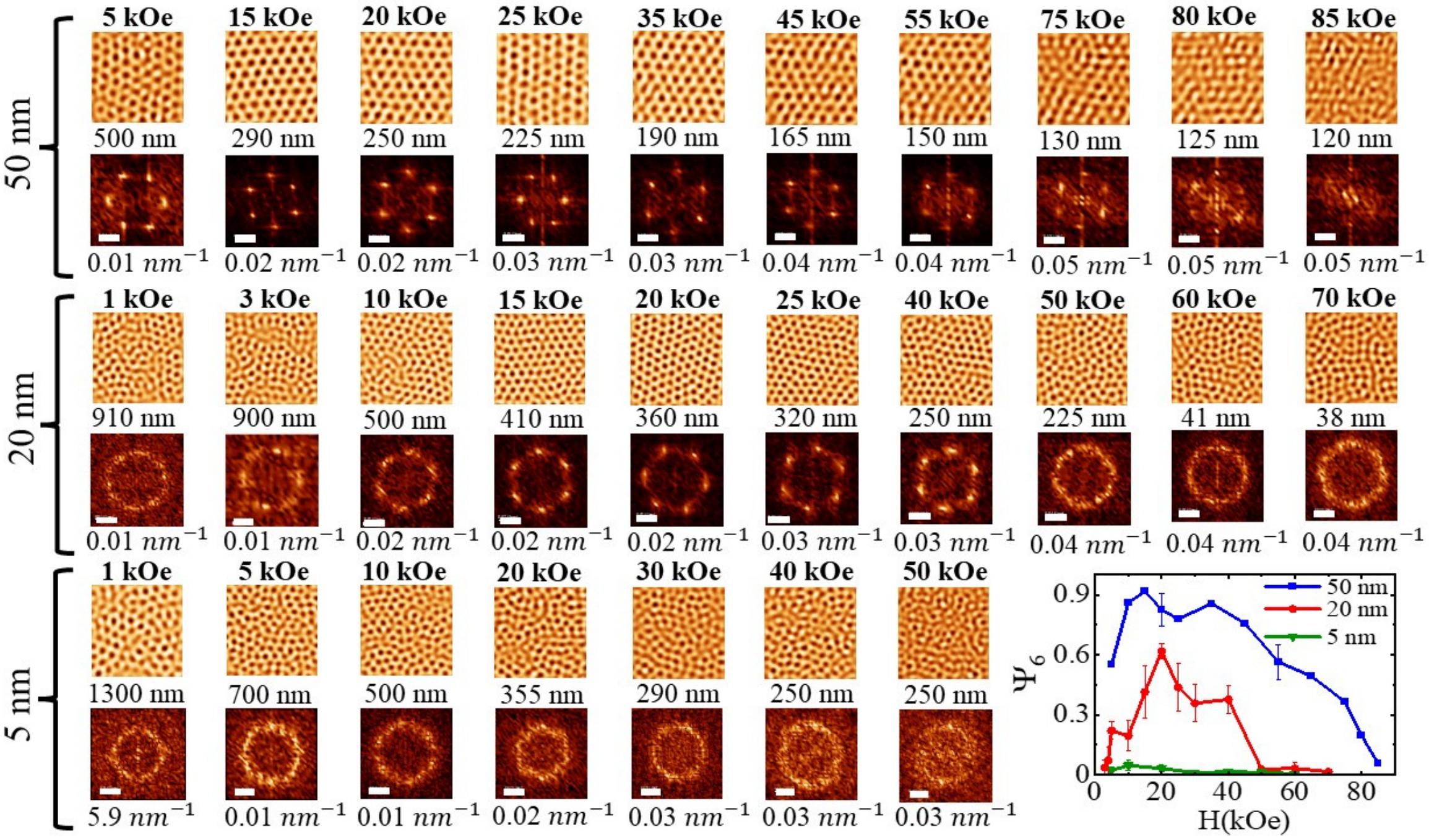

**Figure 1.** Real-space STS conductance maps of vortex configurations (upper panels) and the corresponding two-dimensional Fourier transforms (2DFT) (lower panels) measured at 450 mK for 50 nm, 20 nm, and 5 nm amorphous $Re_6Zr$ films under different magnetic fields. Real space images are filtered as described in ref. 17 to reduce noise. The 2DFT in the 50 nm film exhibits well-defined sixfold Bragg spots over the entire magnetic field range, indicating a hexagonal vortex solid. The 5 nm film shows a diffuse ring at all fields corresponding to a disordered vortex state. In contrast, the 20 nm film displays a re-entrant ordering behavior, evolving from a disordered state at low fields to an ordered hexagonal vortex lattice at intermediate fields before disordering again at higher fields. The image size and reciprocal-space scale bars are shown below each panel. The bottom-right panel shows the magnetic-field dependence of the orientational order parameter $\Psi_6$ extracted from the vortex configurations.

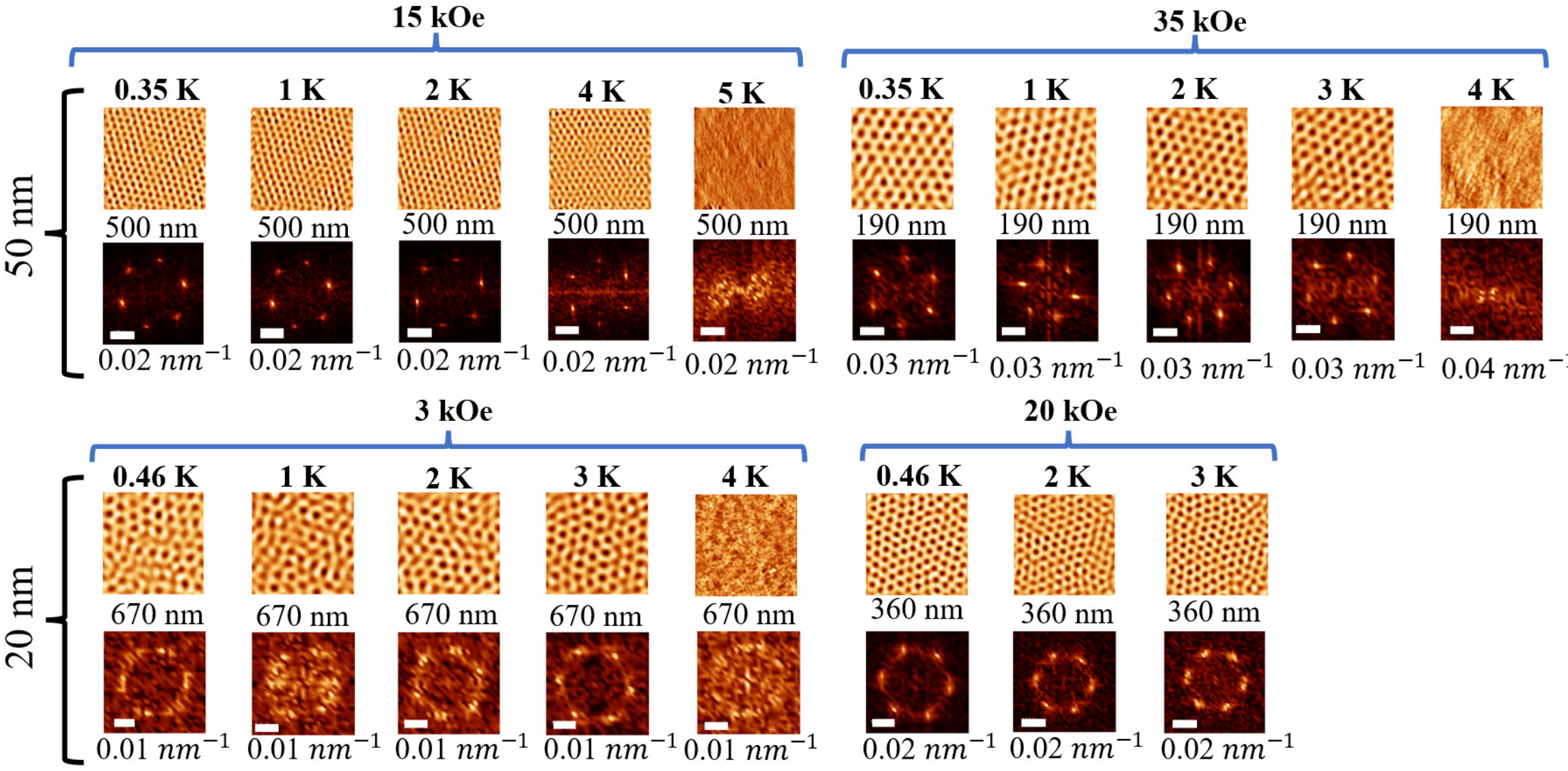

**Figure 2.** Temperature evolution of vortex configurations (upper panels) and corresponding two-dimensional Fourier transforms (2DFT) (lower panels) for the 50 nm and 20 nm amorphous $Re_6Zr$ films at representative magnetic fields obtained from STS conductance maps. Real space images are filtered as described in ref. 17 to reduce noise. The 50 nm film retains sharp sixfold Bragg spots and melts abruptly above a characteristic temperature. In contrast, the 20 nm film exhibits a non-monotonic evolution at low fields, where the vortex lattice evolves from a disordered state at low temperature to an ordered hexagonal configuration at intermediate temperatures before disordering again at higher temperatures. The image size and reciprocal-space scale bars are shown below each panel.

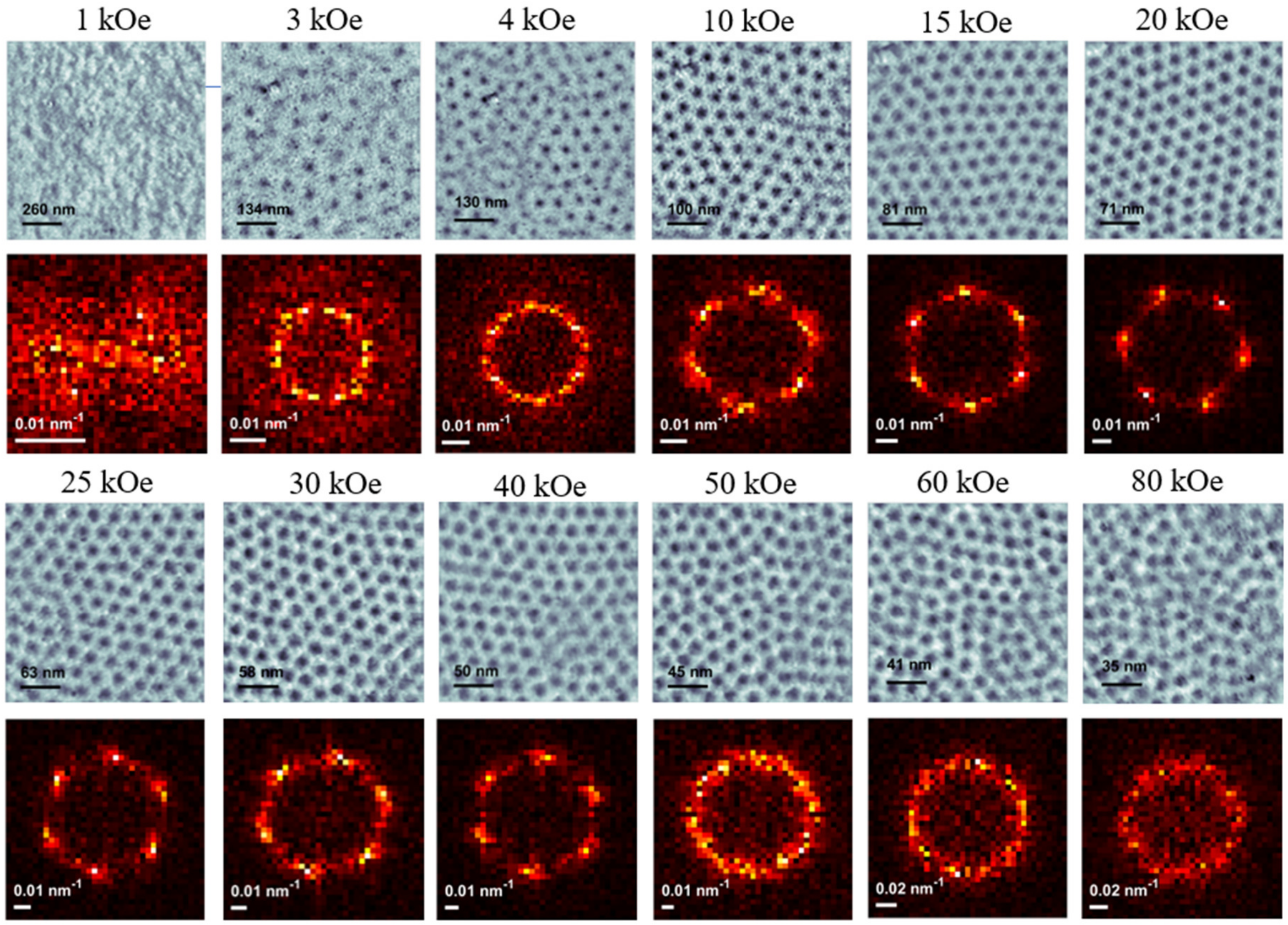

**Figure 3.** Magnetic-field evolution of the summed vortex images and corresponding two-dimensional Fourier transforms (2DFT) for the 20 nm amorphous $Re_6Zr$ film measured at 460 mK. The system evolves from an inhomogeneous vortex liquid at low fields to an ordered hexagonal vortex lattice at intermediate fields, followed by an inhomogeneous vortex liquid at high magnetic fields. The image size scale bars and reciprocal-space scale bars are indicated in the respective panels.

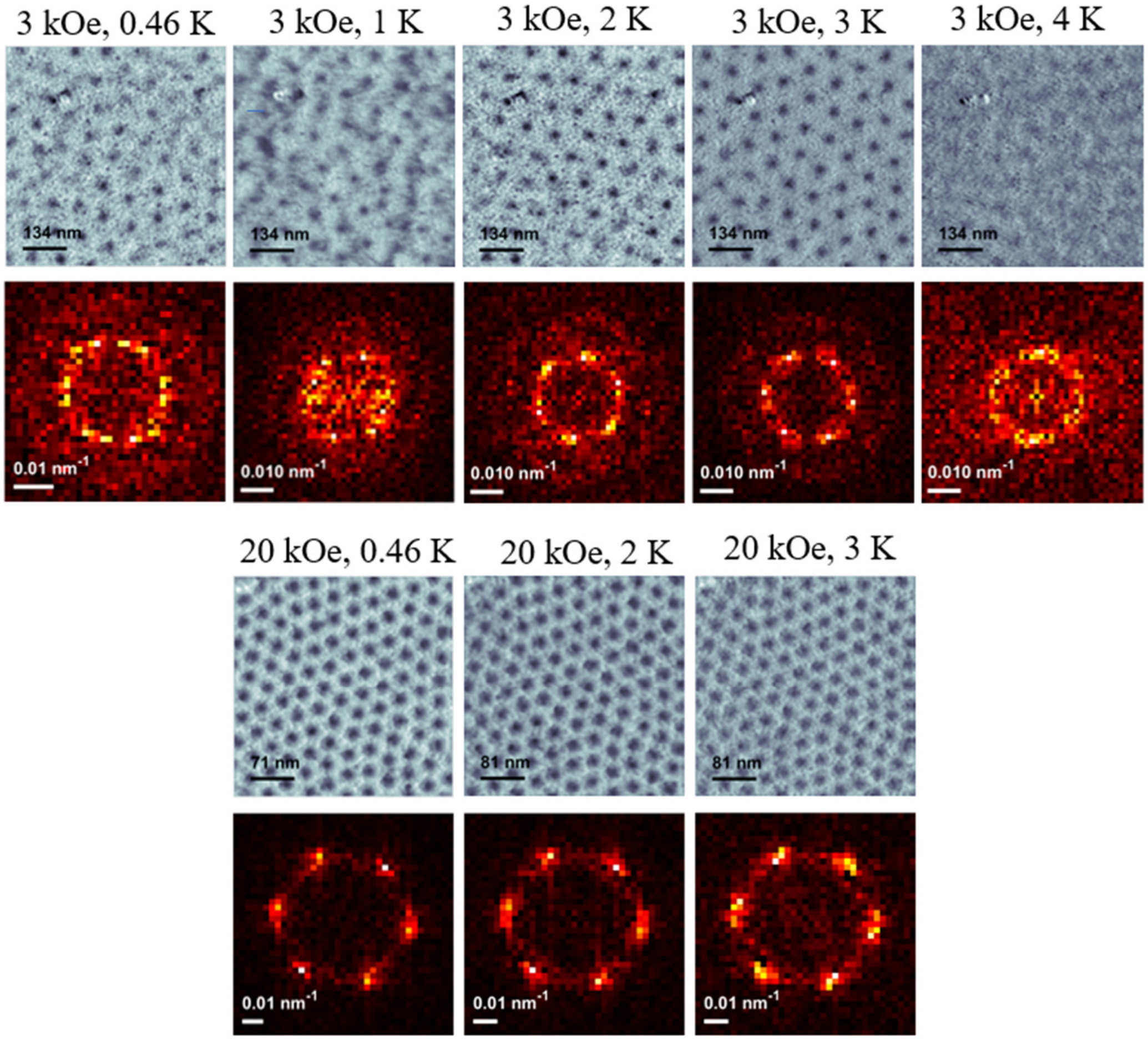

**Figure 4.** Temperature evolution of the summed STM vortex maps and corresponding two-dimensional Fourier transforms for the 20 nm amorphous $Re_6Zr$ film at 3 kOe and 20 kOe. At 3 kOe, the vortex configuration evolves from a disordered state at low temperature to a more ordered hexagonal arrangement at intermediate temperatures before disordering again at higher temperatures, indicating inverse melting behaviour. In contrast, the vortex lattice at 20 kOe remains ordered over the measured temperature range.

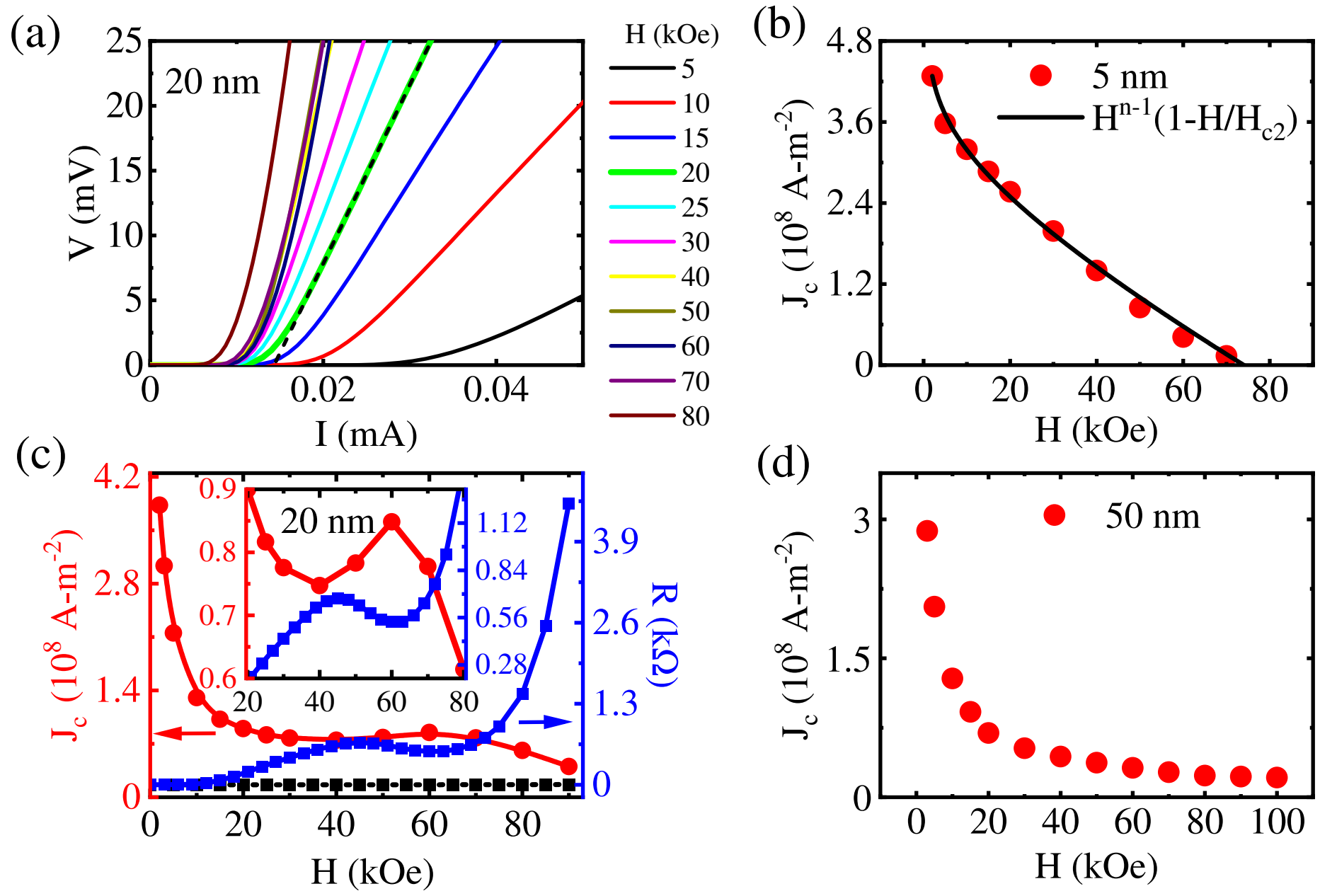

**Figure 5.** Magneto-transport response of amorphous $Re_6Zr$ films measured at 300 mK. (a) Representative current voltage ($I$–$V$) characteristics of the 20 nm film at different magnetic fields. The critical current, $I_c$, is determined by extrapolating the linear flux-flow regime to the current axis, as illustrated by the black dashed line for 20 kOe. (b) Magnetic-field dependence of $J_c$ for the 5 nm film together with a fit to the individual vortex pinning form $J_c \propto H^{n-1}(1 - H/H_0)$, with $n = 0.87$ and $H_0 = 76$ kOe. (c) Magnetic-field dependence of $J_c$ for the 20 nm film along with the magnetoresistance measured using probing currents of $I = 500$ nA (black points) and $I = 15$ μA (blue points). The 20 nm film exhibits a pronounced peak effect in $J_c$, accompanied by a corresponding dip in resistance for $I = 15$ μA, associated with the order-disorder transition of the vortex solid. No corresponding feature is observed in the low-current magnetoresistance measured at 500 nA. The *inset* shows an expanded view close to the peak effect. (d) Magnetic-field dependence of $J_c$ for the 50 nm film, showing a monotonic decrease characteristic of weak collective pinning in an ordered vortex lattice.

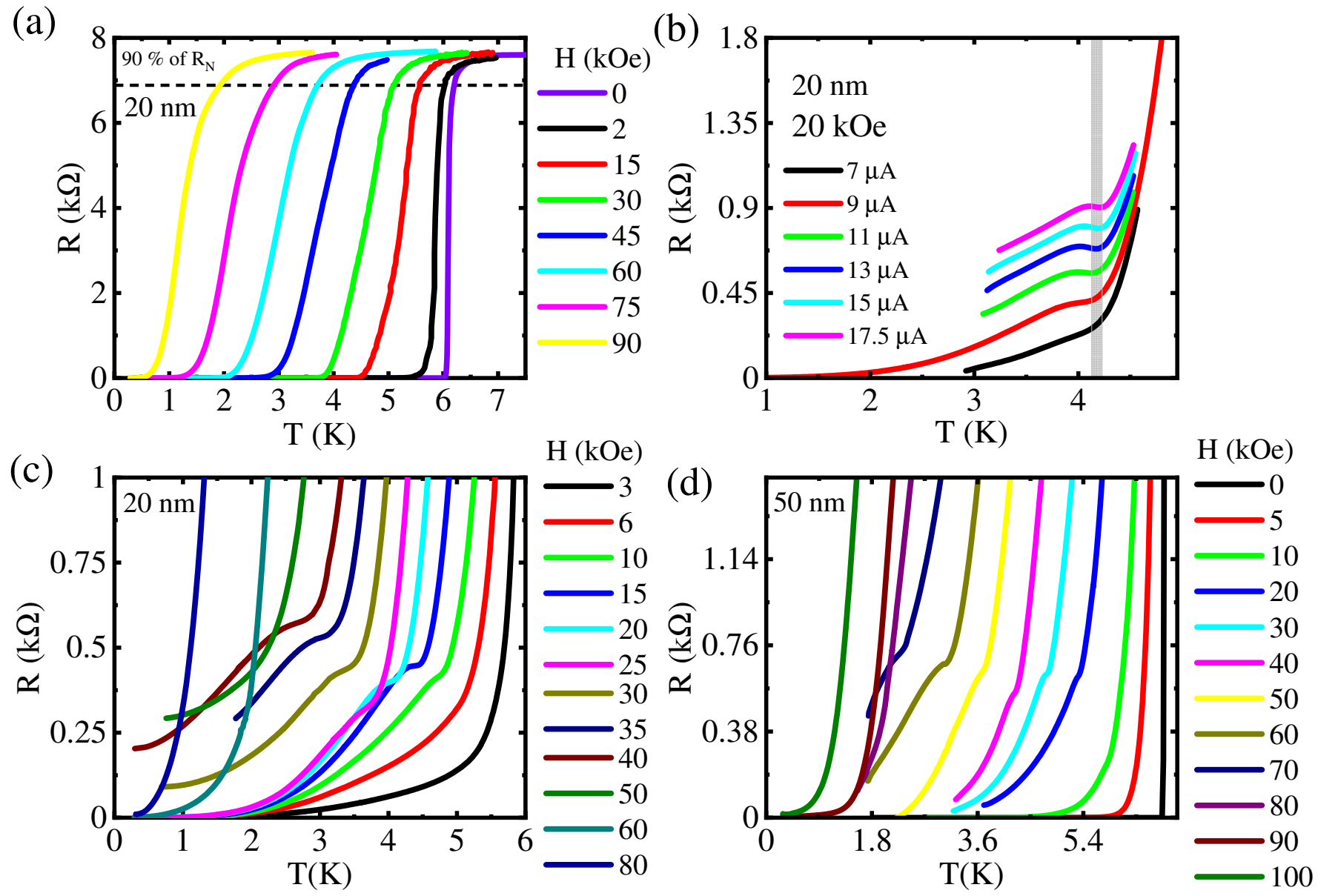

**Figure 6.** Temperature-dependence of magneto-transport measurements in amorphous Re₆Zr films. (a) $R$-$T$ curves for the 20 nm film measured at different magnetic fields with a low probing current ($I = 500$ nA $\ll I_c$); $H_{c2}(T)$is determined from the points where the resistance is 90% of normal state resistance ($R_N$). (b) $R$-$T$ curves for the 20 nm film at 20 kOe measured with different probing currents close to $I_c$, showing the emergence of the peak-effect anomaly as a dip in resistance. The position of the resistance minima varies at most by 90 mK between 9 $\mu A$ (the lowest current where it is observed) and 17.5 $\mu A$ as indicated by the shaded region. (c) Magnetic field evolution of the peak effect in the 20 nm film, measured with currents close to $I_c$ at each field. The peak effect is most prominent between 10–40 kOe, corresponding to the disordering of the vortex solid. (d) $R$-$T$ curves for the 50 nm film measured near $I_c$, displaying peak-effect over a wider magnetic field range.

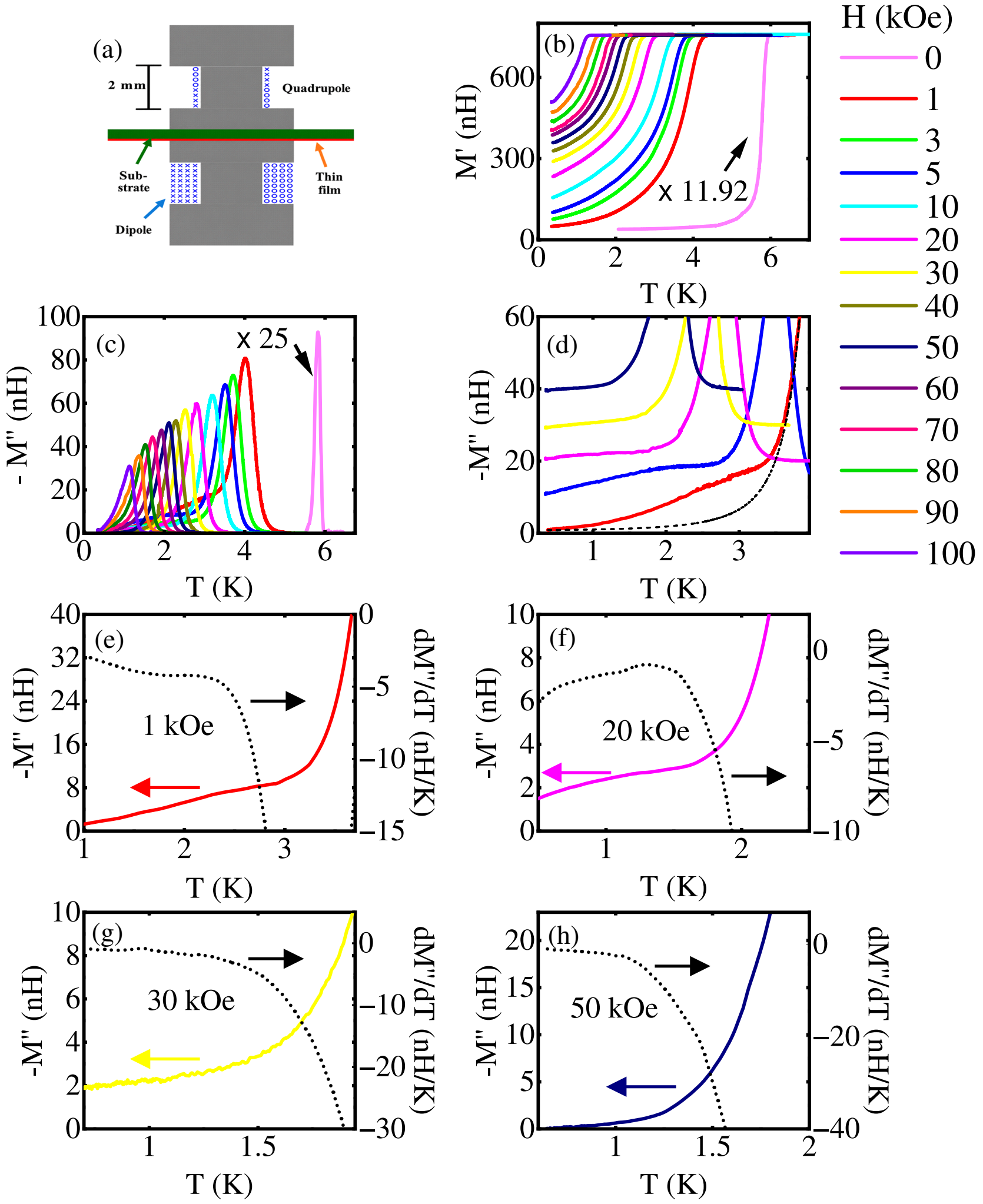

**Figure 7.** The AC magnetic screening response of the 20 nm amorphous $Re_6Zr$ film. (a) Schematic of the two-coil mutual inductance measurement. Temperature dependence of the (b) real $(M')$ and (c) imaginary $(-M'')$ components of the mutual inductance measured at different magnetic fields. While $M'$shows diamagnetic response below $T_c(H)$, the dissipative response $-M''$ exhibits additional anomalies for $H \leq$ 25 kOe. The zero-field data was obtained in a separate cryostat enclosed in a mu-metal shield to eliminate any stray magnetic field. (d) Expanded view of $-M''$ highlighting the shoulder/minimum observed for $H \leq$ 25 kOe; this feature is associated with the inverse melting of the vortex lattice. (e)–(h) Temperature dependence of $-M''$ together with $dM''/dT$ at representative magnetic fields; below 25 kOe the shoulder in $-M''$ appears as a local maximum in the derivative.

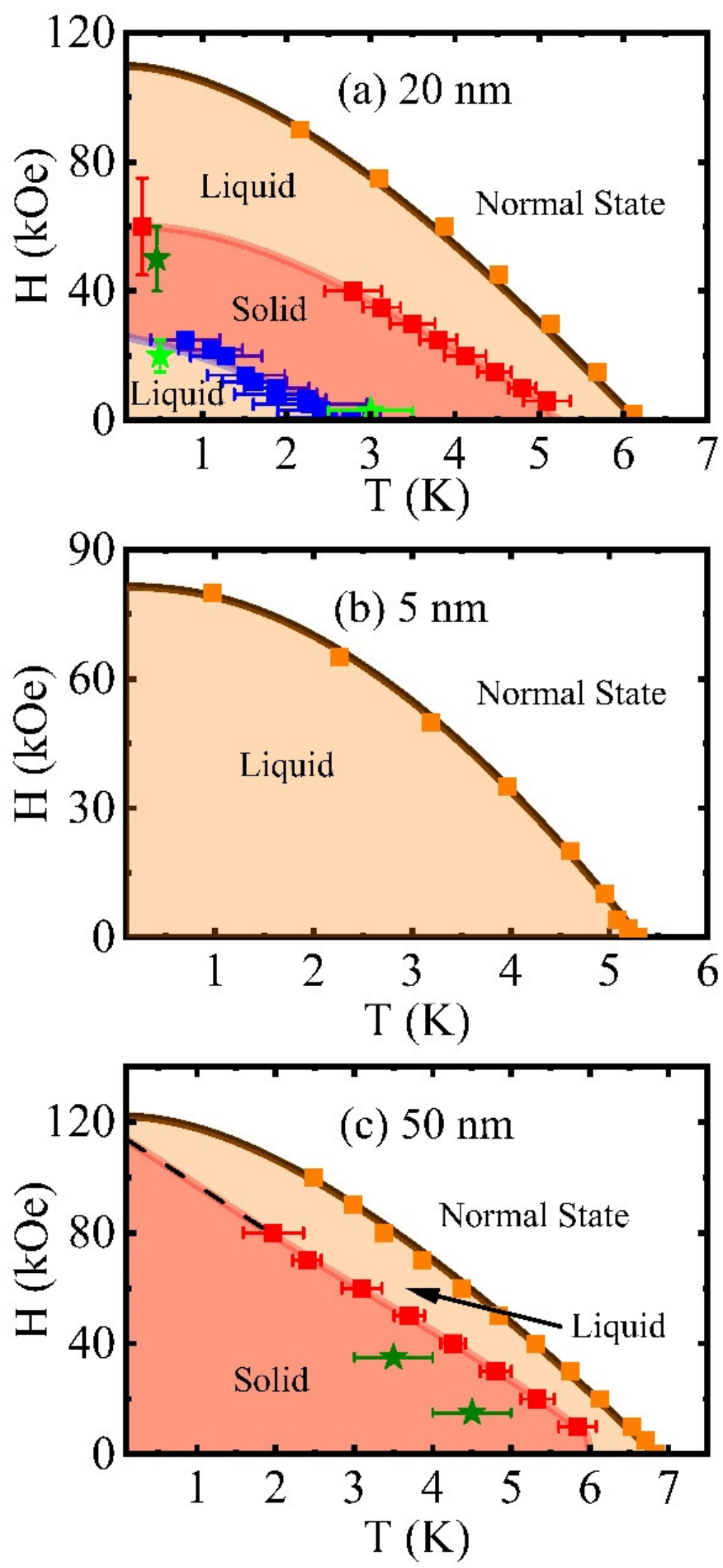

**Figure 8.** Magnetic field–temperature ($H$–$T$) vortex phase diagrams for *a*-ReZr films of different thicknesses: (a) 20 nm, (b) 5 nm, and (c) 50 nm. The orange squares show the upper critical field, $H_{c2}$, obtained from fields where the resistance is 90 % of the normal state value; the solid brown lines are WHH fits to the data. The red squares show the melting line obtained from the loci of the peak effect in transport measurements. The blue squares show the inverse melting line obtained from magnetic shielding measurements. In the 5 nm thick film vortices remain in a liquid state in over the entire superconducting state. In contrast, the 50 nm thick film predominantly exhibits a vortex solid phase, which melts into a vortex liquid close to $H_{c2}$. Only the 20 nm thick film shows both a melting line and an inverse melting line, revealing re-entrant liquid–solid–liquid transformations as a function of magnetic field and temperature. The stars in panel (a) and (c) correspond to the melting and inverse melting points obtained from STS images.

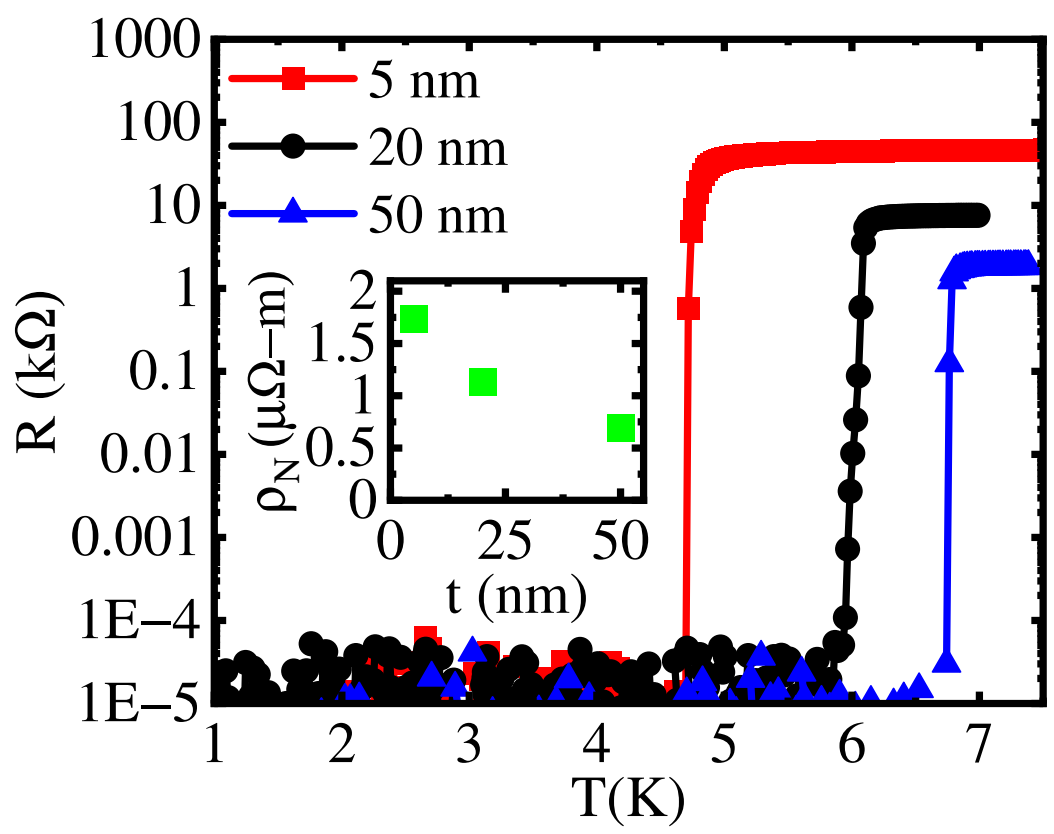

**Figure 9.** Temperature dependence of resistance $R(T)$ for the 5 nm, 20 nm, and 50 nm $Re_6Zr$ Hall bar samples with a channel width of 10 μm. All samples exhibit sharp superconducting transitions with critical temperatures $T_c$ = 4.7 K, 5.8 K, and 6.5 K for the 5 nm, 20 nm, and 50 nm films, respectively. The inset shows the normal state resistivity as a function of film thickness.

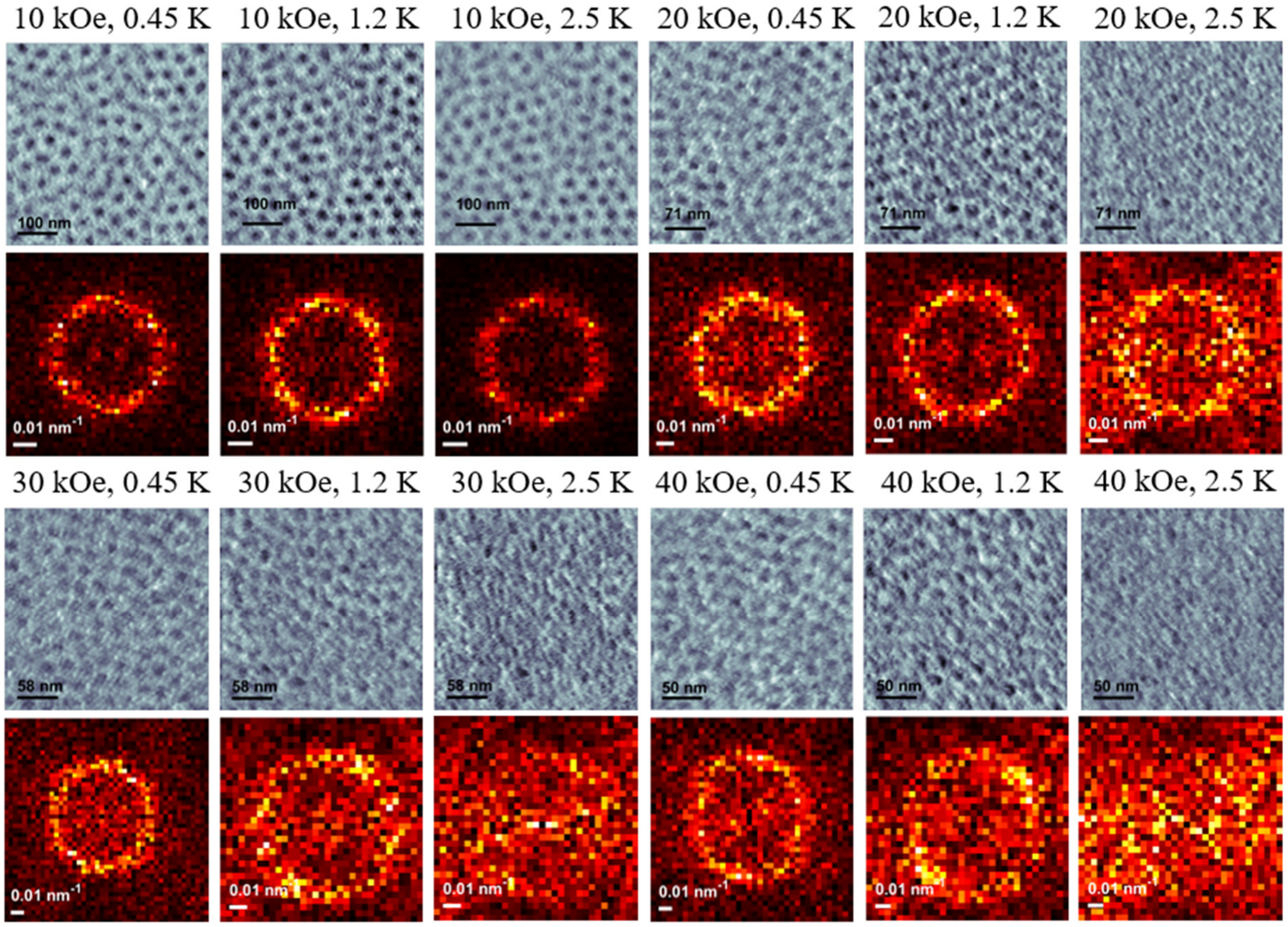

**Figure 10.** Representative summed raw real-space STS conductance maps of vortex configurations (top panels) and their corresponding FFTs (bottom panels) for the 5 nm *a*-ReZr film measured in magnetic fields of 10 kOe, 20 kOe, 30 kOe and 40 kOe and temperatures of 0.45 K, 1.2 K, and 2.5 K. For each field and temperature, 10 consecutive STM images were summed and the 2DFT was calculated from the summed image. The image size scale bars and reciprocal-space scale bars are indicated in the respective panels.

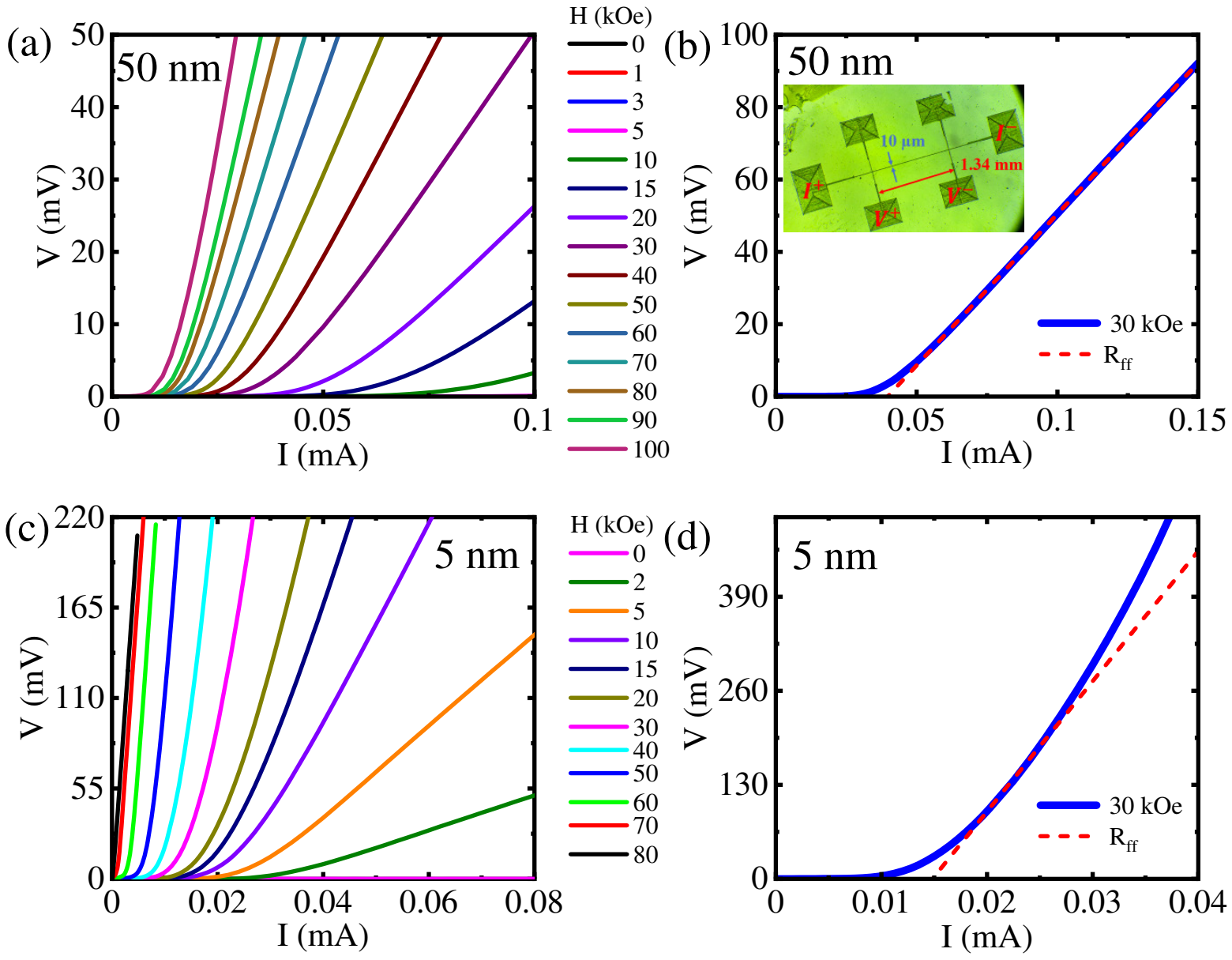

**Figure 11.** (a) and (c) Current–voltage (I–V) characteristics measured at 300 mK for the 50 nm and 5 nm *a*-ReZr samples, respectively, at different applied magnetic fields. (b) and (d) Representative I–V curve (blue line) for the 50 nm and 5 nm sample at 300 mK and 30 kOe. The dashed red line represents a tangent drawn to the region where the slope corresponds to the Bardeen–Stephen flux-flow resistance $R_{ff}$. The critical current $I_c$ is determined from the intercept of this tangent with the current axis. The *inset* of panel (b) shows the geometry of the device on which transport measurements are performed.

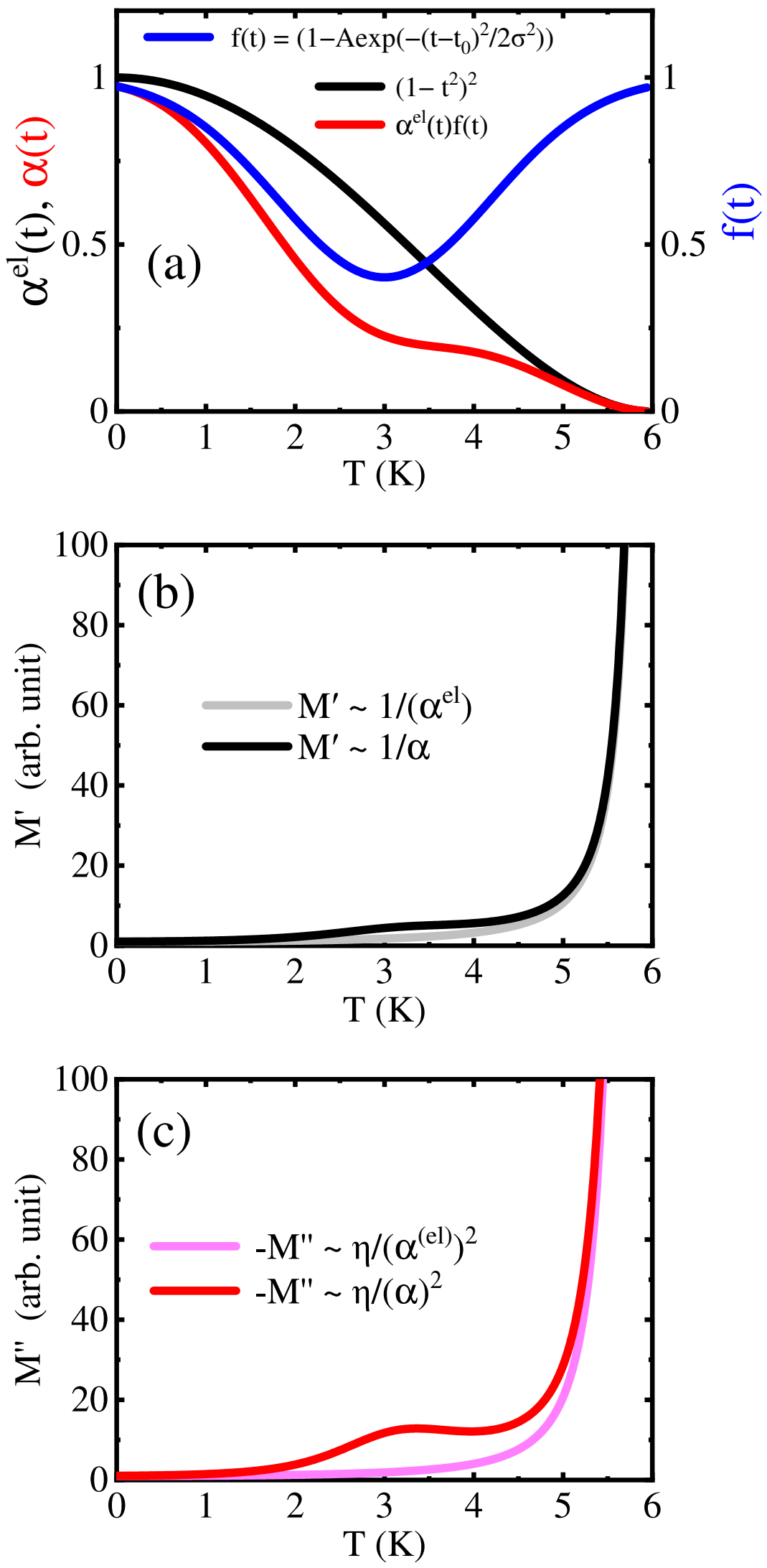

**Figure 12.** (a) Temperature dependence of the elementary pinning contribution $\alpha^{el}(t) \propto (1 - t^2)^2$ (black), $f(t) = 1 - A\exp\left[-(t - t_0)^2/2\sigma^2\right]$ (blue), where A = 0.3, $t_0$ = 0.5 and $\sigma$ = 0.2, and $\alpha(t) = \alpha^{el}(t)f(t)$ (red); here t = $T/T_C$ . Calculated values of (b) $M' \propto 1/\alpha$ and (c) $-M'' \propto \eta/\alpha^2$, with and without collective pinning correction; $\eta \propto (1 - t^2)/(1 + t^2)$ is the viscous drag coefficient. While inverse melting produces only a weak feature in $M'(T)$, the stronger $1/\alpha^2$ dependence enhances the anomaly in $-M''(T)$, resulting in a shoulder-like feature like that observed experimentally.